\documentstyle[12pt]{article}
\def\p{\partial}
\def\t{\tilde}

\def\l{\lambda}
\def\z{\zeta}
\def\s{\sigma}
\def\e{\epsilon}
\def\a{\alpha}
\def\g{\gamma}
\def\b{\beta}
\def\d{\delta}
\def\D{\Delta}
\def\t{\tilde}
\def\O{\Omega}
\def\o{\omega}
\begin{document}

\begin{titlepage}

\begin{flushright}
 FR-THEP/96-10 \\
 hep-th/9606077\\ 
 June 1996
\end{flushright}
\vspace{2.5cm}
\begin{center}
{\Large \bf Symmetries, Currents and Conservation Laws}

\vspace{0.3cm}

{\Large \bf of Self-Dual Gravity}

\vspace{1cm}

{\large  A.D.Popov\footnote{Supported by 
the Alexander von Humboldt Foundation}$^,$\footnote{On 
leave of absence from Bogoliubov Laboratory
of Theoretical Physics, JINR, Dubna, Russia}, M.Bordemann
 and H.R\"omer}\\

\vspace{1cm}

{\it Fakult\"at f\"ur Physik, Universit\"at Freiburg,\\
Hermann-Herder-Str. 3, 79104 Freiburg, Germany}

\vspace{1cm}

{\small popov@phyq1.physik.uni-freiburg.de\\
mbor@phyq1.physik.uni-freiburg.de\\
roemer@phyq1.physik.uni-freiburg.de}     

\vspace{1.5cm}

{\bf Abstract}
\end{center}

We describe  an infinite-dimensional algebra of hidden symmetries for 
the self-dual gravity equations. Besides the known diffeomorphism-type 
symmetries (affine extension of $w_\infty$ algebra), this algebra
contains new hidden symmetries, which are an affine extension of the
Lorentz rotations. The full symmetry algebra has both Kac-Moody
and Virasoro-like generators, whose exponentiation maps solutions of 
the field equations to other solutions. Relations to problems of string 
theories are briefly discussed.

\end{titlepage}

{\bf 1. Introduction}

\vspace{0.2cm}

The purpose of this paper is to describe a new infinite-dimensional
algebra of hidden symmetries of the self-dual Einstein equations on a
metric of signature $(+ + + +)$ or $(+ + - -)$. These equations define
manifolds with self-dual Weyl tensor and vanishing Ricci tensor, which is 
equivalent to the self-duality equations for the Riemann tensor.

{}Four-dimensional self-dual Euclidean backgrounds often arise as the 
internal part of superstrings compactified to six dimensions
in consideration of consistent string propagation (see, e.g., [1]
and references therein). Self-dual gravity configurations also
arise as consistent backgrounds for the $N=2$ closed
string theory [2,3], and the $N=2$ string theory provides a quantization 
of the self-dual gravity model in a space-time with signature (2,2).
Self-dual geometries are also important in compactifications of 
recently proposed 12-dimensional fundamental Y- and F-theories [4].
It is believed that discrete subgroups of the classical symmetry group of 
consistent string backgrounds are symmetries of string theory and that 
these subgroups of a large hidden symmetry group of string theory
become visible for various compactifications [5]. Therefore hidden
symmetries of the self-dual gravity equations are relevant to the 
symmetries of string theories.

The study of these symmetries is important for an understanding of 
non-perturbative properties and quantization of gravity and string
theories. Euclidean solutions of the self-dual gravity equations
(gravitational instantons) give a main contribution to a path integral
of quantum gravity (see, e.g., [6]), and quantization 
of the self-dual gravity model itself may provide useful hints for 
full quantum gravity (see, e.g., [7]).

The self-duality equations on the curvature of a metric in four
dimensions are an important example of a multidimensional integrable 
system,
which can be solved by a twistor geometric construction [8--11]. 
The discussion of hidden symmetries of this model was started 
in the papers [12] on the basis
of Pleba\'nski's equations [13], and has been continued by many authors 
(see, e.g., [14--18]). For the study of hidden symmetries the
reformulation of the self-dual gravity equations as (reduced) 
self-dual Yang-Mills equations
with infinite-dimensional gauge group was very useful [19,16,20]
(see also the clear exposition in [21]). It was shown that the self-dual
gravity
equations are invariant with respect to a group, whose generators form 
the affine Lie algebra $w_\infty\otimes C[ \l , \l^{-1}]$, $\l\in C$,
associated with the Lie algebra $w_\infty$ of area-preserving 
diffeomorphisms of a certain (null) surface [12,14--18].

We shall make a further step in the investigation of hidden symmetries of 
the self-dual gravity (SDG) equations. Our main results are the following:

\begin{itemize}
\item To each Lorentz rotation of the tangent space we associate
      an infinite number of new symmetries of the SDG equations 
      and conserved currents. We show that these symmetries form 
      a Kac-Moody-Virasoro type algebra, in fact the same as the one 
      considered in [22].
      These symmetries underlie the cancellation of almost all amplitudes
      in the theory of $N=2$ closed self-dual strings [2, 3].
\item We define the action of the classical algebra $w_\infty$ on 
      the (conformal)
      tetrad and, using certain operator product expansion type formulae, 
      we 
      present a new derivation of the symmetry algebra 
      $w_\infty\otimes C[\l ]
      \subset w_\infty\otimes C[\l , \l^{-1}]$ of the SDG equations. 
      We also describe the commutation relations between the generators 
      of the `old' and the `new' symmetries.
\item It is well-known that for metrics with `rotational' Killing symmetry
      the SDG equations are reduced to the continual Toda equation 
      ($sl(\infty )$-Toda field equation) [23], and for 
      metrics with `translational' Killing symmetry the SDG equations
     are reduced to the Gibbons-Hawking equations [24] 
    in three dimensions. By reduction of the symmetry algebra of 
     the SDG equations
      we obtain the well-known symmetry algebra $w_\infty$ of the continual
      Toda equation [25] and the symmetry algebra of the Gibbons-Hawking 
      equations, which has not appeared in the literature before.
\item Recently, it was found that the T-duality transformation with respect 
     to the rotational Killing
      vector fields 
     (i.e. those which do not in general preserve the complex structure(s))
     does not preserve the self-duality conditions, that leads
      to apparent violations of the $N=4$ world-sheet supersymmetry [26] 
     (see also  [27]). The
      T-duality transformation with respect to the translational 
      Killing vector fields (i.e. those which preserve the complex 
      structure(s)) preserves the self-duality 
      conditions. We show that the translational vector fields generate
      the Abelian loop group $LU(1)=C^\infty(S^1, U(1))$ of symmetries
      of the SDG equations,
      and the rotational vector fields generate the non-Abelian Virasoro 
      symmetry group Diff($S^1$). This ``non-Abelian nature"
      of the rotational Killing vector fields underlies the nonpreservation 
      of the 
      local realizations of the world-sheet and space-time supersymmetries 
      under the T-duality
      transformation with respect to such Killing vector fields. 
\end{itemize}

In this paper we describe new hidden symmetries of the SDG equations 
omitting direct computations and writing out only the final formulae.

\vspace{1cm}

{\bf 2. Manifest symmetries of self-dual gravity}

\vspace{0.2cm}

Let $M^4$ be a complex four-dimensional manifold with a nondegenerate
complex holomorphic metric $g$. We shall suppose that 
$M^4$ is oriented and denote by $\o$ a complex holomorphic volume four-form.
Consider the infinite-dimensional algebra sdiff$(M^4)$ of volume-preserving 
vector fields on $M^4$. For  $N=N^\mu\p_\mu\in$ 
sdiff$(M^4)$ ($\mu ,\nu,...=1,...,4\ $)
the Lie derivative of $\o$ along $N$ should vanish 
(divergence free vector fields).
Here, and throughout the paper, we use the Einstein summation convention.

Self-dual vacuum (i.e. Ricci flat) metrics may be constructed
as follows [19]: For four pointwise linearly independent vector fields 
$B_\a \in$ sdiff$(M^4)$ let us consider the following equations:
$$
\frac{1}{2}\e_{\a\b}\,^{\g\d} [B_\g , B_\d ]=[B_\a , B_\b ],
\eqno(1)
$$
where $\a ,\b ,...=1,...,4$ are Lorentz indices. If one introduces
the vector fields
$$
V_1=\frac{1}{2}(B_1-iB_2), \ V_{\t 1}=\frac{1}{2}(B_1+iB_2), 
\ V_2=\frac{1}{2}(B_3-iB_4), \ V_{\t 2}=\frac{1}{2}(B_3+iB_4)
,\eqno(2)
$$
then one may rewrite eqs.(1) in the form
$$
[V_{\t 1}, V_{\t 2}]=0, \ [V_{\t 1}, V_1]-[V_{\t 2}, V_2]=0,\ [V_1, V_2]=0
.\eqno(3)
$$
{}Finally, let $f$ be a scalar, a conformal factor, defined by 
$f^2=\o (V_1, V_2, V_{\t 1}, V_{\t 2})$. Then one may define a 
(contravariant) metric 
\newpage
$$
g=f^{-2}(V_1\otimes V_{\t 1} + V_{\t 1}\otimes V_1-
V_2\otimes V_{\t 2} - V_{\t 2}\otimes V_2)\ \Leftrightarrow
\eqno(4a)
$$
$$
g^{\mu\nu}=f^{-2} g^{A\t A}(V_A^\mu V_{\t A}^\nu +V_A^\nu V_{\t A}^\mu ),
\eqno(4b)
$$ 
where $g^{1\t 1}=g^{\t 1 1}=-g^{2\t 2}=-g^{\t 2 2}=1,\ A,B,...=1,2,\ 
\t A, \t B,...=1,2,\ $ and the Riemann tensor of this metric will be 
self-dual. Conversely, every self-dual vacuum metric arises in this way.
{}For proofs and discussions see [19--21]. We call eqs.(3)
(and eqs.(1)) the self-dual gravity (SDG) equations. Notice, that $\{ 
f^{-1}V_{\t A}, f^{-1}V_A\}$ is a null tetrad for the self-dual
vacuum metric (4).

An infinitesimal symmetry transformation of a system of partial
differential equations is a map $\d :\ s\rightarrow \d s$, which to
each solution $s$ of the system assigns a solution $\d s$ of the linearized
(around $s$) form of the system. The linearized form of the system
may be derived by substituting $s+\e\d s$ into the system, and keeping only 
terms of the first order in the parameter $\e$. In particular, for eqs.(1)
we obtain the following equations on $\d B_\a$: 
$$
\e_{\a\b}\,^{\g\s} [B_\g , \d B_\s ]=[B_\a , \d B_\b ] +[\d B_\a , B_\b ]
.\eqno(5)
$$

{}For any two vector fields $M, N$ in the algebra sdiff$(M^4)$ we define 
the transformations of the vector fields $\{ B_\a\}$ as follows:
$$
\d^0_M B_\a := [M, B_\a ]\ \Rightarrow \ 
[\d^0_M, \d^0_N]B_\a =\d^0_{[M,N]} B_\a 
.\eqno(6)
$$
Substituting (6) into (5) and using the Jacobi identities, it is not hard
to show that $\d^0_MB_\a$ satisfy eqs.(5), i.e. $\d^0_M$ is a symmetry 
of eqs.(1).

Let us now consider global (not depending on coordinates) Lorentz 
rotations, which form the algebra $so(4,C)\simeq sl(2, C)\oplus sl(2, C)$, 
with the generators $\{W_i\,_\a^\b \}=\{X_a\,_\a^\b, X_{\hat a}\,_\a^\b\}$:   
$$
[X_a, X_b]= f_{ab}^c X_c, \ [X_a, X_{\hat b}]=0, \ 
[X_{\hat a}, X_{\hat b}]= f_{\hat a\hat b}^{\hat c} X_{\hat c}
,\eqno(7)
$$
where $i, j,...=1,...,6;\ a, b,...=1,2,3;\ \hat a, \hat b,...=1,2,3;
$ and $f_{12}^3=f_{\hat 1\hat 2}^{\hat 3}=-f_{23}^1=$  $
-f_{\hat 2\hat 3}^{\hat 1}=-f_{31}^2=-f_{\hat 3\hat 1}^{\hat 2}=1$ are 
the structure constants of the algebra
$sl(2, C)$. Let us define the following transformations
$\D_{W_i}$ of the vector fields $\{B_\a\}$:
$$
\D_{W_i} B_\a :=W_i\,_\a^\b B_\b \ \Rightarrow \ [\D_{W_i}, \D_{W_j}]B_\a 
= -\D_{[W_i,W_j]} B_\a
.\eqno(8)
$$
One may consider $\{B_\a\}$ as a vector field with extra Lorentz index
$\a$. We write out the explicit formulae for the components
$W_i\,_\a^\b$ of the matrices $W_i$, defining the action of the 
transformations (8) on the vector field with components $\{V_{\t A}, 
V_A\}$ in the null frame:
$$
\D_{X_1}V_{\t 1} = - \frac{i}{2}V_{\t 2},\ 
\D_{X_1}V_{\t 2} =  \frac{i}{2}V_{\t 1},\ 
\D_{X_1}V_{ 1} =  \frac{i}{2}V_{ 2}, \
\D_{X_1}V_{ 2} = - \frac{i}{2}V_{1},
\eqno(9a)
$$
$$
\D_{X_2}V_{\t 1} =  \frac{1}{2}V_{\t 2},\ 
\D_{X_2}V_{\t 2} =  \frac{1}{2}V_{\t 1},\ 
\D_{X_2}V_{ 1} =  \frac{1}{2}V_{ 2}, \
\D_{X_2}V_{ 2} = \frac{1}{2}V_{1},
\eqno(9b)
$$
$$
\D_{X_3}V_{\t 1} = - \frac{i}{2}V_{\t 1},\ 
\D_{X_3}V_{\t 2} =  \frac{i}{2}V_{\t 1},\ 
\D_{X_3}V_{ 1} =  \frac{i}{2}V_{ 1}, \
\D_{X_3}V_{ 2} = - \frac{i}{2}V_{2},
\eqno(9c)
$$
\newpage
$$
\D_{X_{\hat 1}}V_{\t 1} = - \frac{i}{2}V_{ 2},\ 
\D_{X_{\hat 1}}V_{\t 2} = - \frac{i}{2}V_{ 1},\ 
\D_{X_{\hat 1}}V_{ 1} =  \frac{i}{2}V_{\t 2}, \
\D_{X_{\hat 1}}V_{ 2} =  \frac{i}{2}V_{\t 1},
\eqno(10a)
$$
$$
\D_{X_{\hat 2}}V_{\t 1} =  \frac{1}{2}V_{2},\ 
\D_{X_{\hat 2}}V_{\t 2} =  \frac{1}{2}V_{1},\ 
\D_{X_{\hat 2}}V_{ 1} =  \frac{1}{2}V_{\t 2}, \
\D_{X_{\hat 2}}V_{ 2} = \frac{1}{2}V_{\t 1},
\eqno(10b)
$$
$$
\D_{X_{\hat 3}}V_{\t 1} = - \frac{i}{2}V_{\t 1},\ 
\D_{X_{\hat 3}}V_{\t 2} = - \frac{i}{2}V_{\t 2},\ 
\D_{X_{\hat 3}}V_{ 1} =  \frac{i}{2}V_{ 1}, \
\D_{X_{\hat 3}}V_{ 2} = \frac{i}{2}V_{2}
.\eqno(10c)
$$
It is obvious that
$$
[\d^0_M, \D_{W_i}] B_\a =0
,\eqno(11)
$$
i.e. the transformations (6) and (9), (10) commute.

The symmetries under the transformations (6) in the group SDiff$(M^4)$ are
gauge symmetries, and we may use them for the partial fixing of 
a coordinate system. Namely, from eqs.(3) it follows that one can always 
introduce coordinates $(y, z, \t y, \t z)$ so that $V_{\t 1}$ and
$V_{\t 2}$ become coordinate derivatives (Frobenius theorem), i.e.
$V_{\t A} =\p_{\t A}$,
where $\p_{\t 1}\equiv \p_{\t y}, \ \p_{\t 2}\equiv \p_{\t z}.$
Then $[V_{\t 1}, V_{\t 2}]\equiv 0$,
and the SDG eqs.(3) are reduced to  
$$
g^{\t AA}\p_{\t A} V_A =0\ \Leftrightarrow\ \p_{\t 1} V_1- \p_{\t 2} V_2=0
,\eqno(12a)
$$
$$
\e^{AB}V_A V_B=0\ \Leftrightarrow \ [V_1, V_2]=0,
\eqno(12b)
$$
where $\e^{12}=-\e^{21}=1,$ and we have used the fact that 
$[\p_{\t A}, K]=\p_{\t A}K$ for any vector field $K$.

\vspace{0.3cm}

{\bf Remark.} If we put $\t y=\bar y,\ \t z=\bar z$ (where $\bar y$ and
$\bar z$ are complex conjugate to $y$ and $z$), then the solutions of 
eqs.(12) will define a tetrad on a real self-dual manifold with metric 
(4) of signature $(2,2)$. If we put $\t y=\bar y,\ \t z=-\bar z$, 
then solutions of eqs.(12) will define a tetrad on a real self-dual 
manifold with metric (4) of signature $(4,0)$ (hyper-K\"ahler manifolds).

\vspace{0.3cm}
 
The vector fields $\{V_A\}$ from (12) may be parametrized by a
scalar function (the only degree of freedom of self-dual metrics)
in a different way, and then
eqs.(12) will be reduced to different  nonlinear equations on the scalar 
function (see [13--21]). For example, if we choose 
$$
V_1 =\O_{2\t 2}\p_1 - \O_{1\t 2}\p_2, \
V_2 =\O_{2\t 1}\p_1 - \O_{1\t 1}\p_2, \
\O_{A\t A}\equiv \p_A\p_{\t A}\O ,\
\p_1\equiv \p_y, \ \p_2\equiv \p_z,
\eqno(13a)
$$
then eqs.(12) are reduced to Pleba\'nski's first heavenly equation [13]:
$$
\O_{1\t 2}\O_{2\t 1}-\O_{1\t 1}\O_{2\t 2}=1
.\eqno(13b)
$$
We shall not perform these reductions, because eqs.(12) are more 
fundamental
than various scalar equations [13,16--18], obtained from (12) and 
carrying information about different parametrization of the vector fields 
$\{V_A\}$.

It is obvious that eqs.(12), derived from (3) by partial fixing of the 
coordinate
system (in which $ V_{\t A}=\p_{\t A}$ should not change), will be not
invariant under all the transformations from (6) and (8). The discussion of
residual gauge invariance and of hidden symmetries will be the topic 
of the following Sections.

\newpage

{\bf 3. Affine extension of the $w_\infty \simeq$ sdiff$(\Sigma^2)$ algebra}

\vspace{0.2cm}

It is easy to see that the symmetries of eqs.(12) have to satisfy the 
equations:
$$
\d V_{\t A}\equiv \d \p_{\t A}=0,
\eqno(14a)
$$
$$
\p_{\t 1}\d V_1- \p_{\t 2}\d V_2=0
,\eqno(14b)
$$
$$
[V_1, \d V_2] + [\d V_1, V_2] =0
.\eqno(14c)
$$
As to the transformations (6) from the algebra sdiff$(M^4)$, 
it is evident that 
eqs.(12) will be invariant only under the subalgebra sdiff$(\Sigma^2)
\subset\ $sdiff$(M^4)$ of those vector fields $M, N,...,$ which satisfy
$$
\d^0_M \p_{\t A}:=[\psi^0_M, \p_{\t A}]=0,\
\d^0_M V_{ A}:=[\psi^0_M, V_{A}]\ \Rightarrow 
\eqno(15a)
$$
$$
[\d^0_M, \d^0_N]\p_{\t A} =\d^0_{[M,N]} \p_{\t A}=0,\
[\d^0_M, \d^0_N]V_A =\d^0_{[M,N]} V_A 
.\eqno(15b)
$$
Here we have denoted by $\Sigma^2$ the isotropic
two-dimensional surfaces, parametrized by the coordinates $\{y, z\}$,
and $\psi^0_M:=M$.

It is not difficult to show that the transformations (15a) satisfy
eqs.(14) and from eq.(14b) it follows that $\{\d^0_MV_1, \d^0_MV_2\}$
are two components of the conserved current $\d^0_MV_A$.
{}From (14b) it also follows that there exists a vector field $\psi^1_M$
such that 
$$
\d^0_M V_{ A}\equiv [\psi^0_M, V_{A}] = \e^{\t B}_A\p_{\t B}\psi^1_M,
\eqno(16)
$$
where $\e^{\t B}_A = g^{\t BB}\e_{BA},\ \e_{12}=-\e_{21}=1 \Rightarrow 
\e^{\t 2}_1=\e^{\t 1}_2=1.$
Using $\psi^1_M$, we introduce the transformation $\d^1_M$ by
the formulae:
$$
\d^1_M \p_{\t A}:=0,\
\d^1_M V_{ A}:=[\psi^1_M, V_{A}]
.\eqno(17)
$$

It is not hard to verify that by virtue of eqs.(12), 
$\d^1_M V_{ A}$ satisfies eqs.(14). Therefore, $\d^1_M V_{ A}$
is also a conserved current. Now we may use a standard inductive
procedure that was used, for example, for the construction of
(nonlocal) currents of the chiral fields model [28,29]. 
Namely, let us suppose that we have constructed $\d^n_M$ such that
$$
\d^n_M \p_{\t A}:=0,\
\d^n_M V_{ A}:=[\psi^n_M, V_{A}],\ n\ge 1
.\eqno(18)
$$
Assuming that the current $\d^n_M V_{ A}$ is conserved implies that 
there exists a vector field $\psi^{n+1}_M$ such that
$$
[\psi^n_M, V_{A}] = \e^{\t B}_A\p_{\t B}\psi^{n+1}_M
.\eqno(19)
$$
Using this we shall show that the $(n+1)$-th current
$\d^{n+1}_M V_{ A}:=[\psi^{n+1}_M, V_{A}]$ is conserved, which will 
complete the induction:
$$
\p_{\t 1}\d^{n+1}_M V_1- \p_{\t 2}\d^{n+1}_M V_2=
[\p_{\t 1}\psi^{n+1}_M, V_1] - [\p_{\t 2}\psi^{n+1}_M, V_2]=
$$
$$
=[[\psi^n_M, V_{2}], V_1]+[[V_1, \psi^n_M], V_2]=[[V_1, V_2], \psi^n_M] =0,
\eqno(20a)
$$
$$
[\d^{n+1}_M V_1, V_2] + [V_1, \d^{n+1}_M V_2] =
[[\psi^{n+1}_M, V_1], V_2] + [V_1,[ \psi^{n+1}_M, V_2]] =
[\psi^{n+1}_M,[ V_1, V_2]]=0
.\eqno(20b)
$$
Thus, for any $n\ge 1$ we construct a vector field $\psi^n_M$ and a 
conserved current $\d^{n}_M V_A$, starting from $\psi^0_M:=M$ and
$\d^0_M V_A:=[M, V_A]$.

\vspace{0.3cm}

{\bf Remark.} Using (4), (15), (18) and (19), 
one may show by direct calculations that
$\d^0_M g^{\mu\nu}={\cal L}_{\psi^0_M}g^{\mu\nu}$,
but for $n\ge 1\ $ $\d^n_M g^{\mu\nu}\ne{\cal L}_{\psi^n_M}g^{\mu\nu}$,
where ${\cal L}_{\psi^n_M}$ is a Lie derivative along the vector
field $\psi^n_M$. This means that $\d^0_M$ is a gauge symmetry
(an infinitesimal diffeomorphism), and  $\d^n_M$ with $n\ge 1$ is 
not a gauge symmetry.

\vspace{0.3cm}

Having an infinite number of vector fields 
$\psi^n_M$ on $M^4$, one can introduce the vector field
$\psi_M(y,z,\t y, \t z, \l ):= 
\sum^\infty_{n=0}\l^n\psi^n_M(y,z,\t y, \t z),$
depending on the complex parameter $\l\in C$. Then the infinite number of 
equations (19) ({\it recurrence relations}) may be rewritten
as two linear equations on $\psi_M(\l )$:
$$
\begin{array}{c}
\p_{\t 1}\psi_M + \l [V_2, \psi_M]=0
\\
\p_{\t 2}\psi_M + \l [V_1, \psi_M]=0
\end{array}
\Longleftrightarrow
\ \p_{\t A}\psi_M + \l\e^{B}_{\t A} [V_B, \psi_M]=0,
\eqno(21)
$$
where $\e^{A}_{\t B} = g^{A\t A}\e_{\t A\t B}, 
\ \e_{\t 1\t 2}= -\e_{\t 2\t 1}=1\ \Rightarrow \
\e^{ 2}_{\t 1}=\e^{ 1}_{\t 2}=1$.

\vspace{0.3cm}

{\bf Remark}. Equations (21) can be considered as a linear system
(Lax pair) for the SDG equations (12), because eqs.(12)  are the
compatibility conditions of eqs.(21). As a `canonical' vector field 
one may choose, e.g., $\p_A$ and consider the linear equations
(21) on $\psi_{\p_A}$.

\vspace{0.3cm}

Instead of an infinite number of symmetry generators $\d^n_M$ one may 
introduce
the generator $\d_M(\l ):= \sum^\infty_{n=0}\l^n\d^n_M,\ $
depending on a complex `spectral' parameter $\l\in C$.
It is evident that $\d^n_M =(2\pi i)^{-1} \oint_{C'}d\l\
\l^{-n-1}\d_M(\l ),\ $ where $C'$ is a contour in the $\l$-plane
about the origin. Using $\psi_M(\l )$ and $\d_M(\l )$, formulae
(15a) and (18) may be rewritten in the form of 
a one-parameter family of infinitesimal transformations  
$$
\d_M(\l ) V_{\t A}:=0,\
\d_M(\l ) V_{ A}:=[\psi_M(\l ), V_{A}]
,\eqno(22)
$$
which are  symmetries of eqs.(12) for each $M\in\ $sdiff$(\Sigma^2)$.

Now we are interested in the algebraic properties of the symmetries (22).
It is not difficult to show that
$$
\d_M(\l ) \d_N(\z )V_{ A}:=\d_M(\l )(V_A + \d_N(\z )V_{ A})-\d_M(\l )V_A =
$$
$$
=[\psi_M(\l ) + \d_N(\z )\psi_M(\l ), V_A + \d_N(\z )V_{ A}]-\d_M(\l )V_A
$$
$$
\cong [\d_N(\z )\psi_M(\l ), V_A] + [\psi_M(\l ), \d_N(\z )V_A]
,\eqno(23a)
$$
$$
 \d_N(\z )\d_M(\l )V_A=
[\d_M(\l )\psi_N(\z ), V_A]+[\psi_N(\z ), \d_M(\l )V_A].
\eqno(23b)$$
Then the commutator of two symmetries is equal to
$$
[\d_M(\l ), \d_N(\z )]V_{ A}= [\d_N(\z )\psi_M(\l )-
\d_M(\l )\psi_N(\z )+[\psi_M(\l ), \psi_N(\z )], V_A]
.\eqno(24)
$$
Accordingly, for the variation $\d_N(\z )\psi_M(\l )$ we have
the following equations
$$
\p_{\t A}\d_N(\z )\psi_M(\l )+ \l \e^B_{\t A}[V_B,\d_N(\z )\psi_M(\l )]=
\l \e^B_{\t A}[\psi_M(\l ), \d_N(\z )V_B]
,\eqno(25)
$$
the solutions of which have the form (cf. [29]):
$$
\d_N(\z )\psi_M(\l )=\frac{\z }{\l -\z}(\psi_{[M,N]}(\z )-
[\psi_M(\l ), \psi_N(\z )])\ \Rightarrow
$$
$$
\d_M(\l )\psi_N(\z )=\frac{\l }{\l -\z}(\psi_{[M,N]}(\l )-
[\psi_M(\l ), \psi_N(\z )]).
\eqno(26)
$$
Substituting (26) into (24), we obtain 
$$
[\d_M(\l ), \d_N(\z )]=
\frac{1}{\l -\z}(\l\d_{[M,N]}(\l )- \z\d_{[M,N]}(\z ))\
\Rightarrow\
[\d_M^m, \d_N^n]=\d_{[M,N]}^{m+n},\ m,n\ge 0
,\eqno(27)
$$
when we consider the action on $V_A$ and $V_{\t A}$. 
The algebra (27) is the affine extension sdiff$(\Sigma^2)\otimes C[\l ]$ 
of the algebra sdiff$(\Sigma^2)$
of area-preserving diffeomorphisms. Formulae (27) give us
commutators between half of the generators of the affine
Lie algebra sdiff$(\Sigma^2)\otimes C[\l , \l^{-1}]$.

\vspace{0.5cm}

{\bf Remarks}. 
\begin{enumerate}
\item The described algebra sdiff$(\Sigma^2)\otimes C[\l ]$ of
      symmetries of the SDG equations (12) is known. But the
      formulae (15)--(27),
      describing the action of these symmetries on the (conformal)
      tetrad $\{V_{\t A}, V_A\}$, are new. These 
      formulae may be useful for applications.
\item In the described algebra there is an Abelian subalgebra with
      generators $\{\d^n_{\p_A}\}$, where
      $\{\p_A\}  = \{\p_y, \p_z\}. $ In the usual way (see, e.g., [30]),
       one can associate to this algebra of Abelian symmetries the 
      hierarchy of the SDG equations (cf. [15,17] for other approaches).
\item We restrict our attention to the subalgebra 
      sdiff$(\Sigma^2)\otimes C[\l ]$
      of the symmetry algebra sdiff$(\Sigma^2)\otimes C[\l , \l^{-1}]$.
      The rest will be obtained if we choose the coordinates 
      in such a way that $V_A$
      will be coordinate derivatives (in Sec.2 and Sec.3 $\ V_{\t A}$ 
      were the coordinate derivatives) and consider symmetries of eqs.(3) 
      after    this `dual' partial fixing of coordinates.
\end{enumerate}

\vspace{1cm}

{\bf 4. Hidden symmetries from Lorentz rotations}

\vspace{0.2cm}

The explicit form of the infinitesimal transformations of the vector fields 
$\{V_{\t A}, V_A\}$ under the action of the Lorentz group
$SO(4, C)$ was written out in (9) and (10). From (9), (10) 
one can see that
$\D_{W_i}\p_{\t A}\ne 0$, i.e. these transformations, being
the symmetry of eqs.(3), are not the symmetry of eqs.(12).   In other
words, the transformations (9) and (10) do not preserve the chosen gauge. 
Nevertheless the Lorentz
symmetry can be restored by compensating transformations from the 
diffeomorphism group SDiff$(M^4).$  The point is that from formulae (9), 
(10) it follows that  
$$
\p_{\t 1}(\D_{W_i}\p_{\t 2})-\p_{\t 2}(\D_{W_i}\p_{\t 1})=0
,\eqno(28)
$$
for any $W_i\in so(4, C).$  This means that there exist vector 
fields $\{\psi^0_{W_i}\} = \{
\psi^0_{X_a}, \psi^0_{X_{\hat a}}\}$ such that
$$
\D_{W_i}\p_{\t 1}=\p_{\t 1}\psi^0_{W_i}, \
\D_{W_i}\p_{\t 2}=\p_{\t 2}\psi^0_{W_i}
,\eqno(29)
$$
and one can define the transformation
$$
\d^0_{W_i}\p_{\t A}:=\D_{W_i}\p_{\t A}+[\psi^0_{W_i}, \p_{\t A}]=0,
\eqno(30a)
$$
$$
\d^0_{W_i}V_{A}:=\D_{W_i}V_{A}+[\psi^0_{W_i}, V_{A}],
\eqno(30b)
$$
satisfying the following commutator relations
$$
[\d^0_{W_i}, \d^0_{W_j}] V_{\t A}=\d^0_{[W_i,W_j]}V_{\t A}=0,\
[\d^0_{W_i}, \d^0_{W_j}] V_{A}=\d^0_{[W_i,W_j]}V_{A}
.\eqno(31)
$$
Notice that the equality to zero in (30a) follows from the definition (29)
of the vector fields $\{\psi^0_{W_i}\}$.

\vspace{0.3cm}

{\bf Remark.} Using eqs. (4) and (30), one can 
show by direct computation
that $\d^0_{W_i}$ acts on $g^{\mu\nu}$ as a Lie derivative,
i.e. $\d^0_{W_i}g^{\mu\nu}={\cal L}_{\psi^0_{W_i}}g^{\mu\nu}. $
Therefore, if one defines the action of $\d^0_{W_i}$ on the
vector fields $\psi_M$ from the linear system (21), then one can 
develop the method of reduction for the SDG equations (12) and   
the linear system (21) for them, analogous to the method developed
for the self-dual Yang-Mills model [31,32].

\vspace{0.3cm}

It is not difficult to show that (30) satisfy eqs.(14). So, $\d^0_{W_i}V_A$
is a conserved current and from (14b) it follows that there exists a 
vector field $\psi^1_{W_i}$ such that 
$$
\d^0_{W_i}V_A\equiv \D_{W_i}V_A + [\psi^0_{W_i}, V_A]=
\e^{\t B}_A\p_{\t B}\psi^1_{W_i}.
\eqno(32)
$$
Let us define in full analogy with Sec.3  the transformations
$$
\d^1_{W_i}\p_{\t A}:=0, \
\d^1_{W_i}V_A:=[\psi^1_{W_i}, V_A]
.\eqno(33)
$$
One can verify that (33) is a symmetry of 
eqs.(12). Now with the help of the inductive procedure, identical
to the one described in Sec.3, it is not difficult to show that 
the transformations
$$
\d^{n+1}_{W_i}\p_{\t A}:=0, \quad
\d^{n+1}_{W_i}V_A:=[\psi^{n+1}_{W_i}, V_A]
\eqno(34)
$$
are symmetries of eqs.(12), if 
$$
\d^n_{W_i}V_A\equiv [\psi^n_{W_i}, V_A]=\e^{\t B}_A\p_{\t B}
\psi^{n+1}_{W_i},\ n\ge 1,
\eqno(35)
$$
is a conserved current.

One may introduce the generating vector field 
$\psi_{W_i}(y,z,\t y, \t z, \z ):=$ $\sum^\infty_{n=0}\z^n \psi^n_{W_i}
(y,z,\t y, \t z),$ $ \ \z\in C.$  Then the recurrence relations (35) can be 
collected into the following two linear equations
$$
[\p_{\t A}+\z\e^B_{\t A}V_B, \psi_{W_i}(\z )]=
\D_{W_i}\p_{\t A}+\z \e^B_{\t A}\D_{W_i}V_B.
\eqno(36)
$$
Analogously, introducing 
$\d_{W_i}(\z ):=\sum^\infty_{n=0}\z^n \d^n_{W_i}$, we obtain 
a one-parameter family of infinitesimal transformations 
$$
\d_{W_i}(\z )\p_{\t A}:=0,\ 
\d_{W_i}(\z )V_{ A}:=[\psi_{W_i}(\z ), V_A]+\D_{W_i}V_A .
\eqno(37)
$$
{}For each $W_i\in so(4, C)$ these transformations are new `hidden 
symmetries' of the SDG equations (12).

After some calculations we have the following expression for the commutator 
of two symmetries
$$
[\d_{W_i}(\l ), \d_{W_j}(\z )] V_A=
\e^{\t B}_A\p_{\t B} 
\{\frac{1}{\l}\d_{W_j}(\z )\psi_{W_i}(\l )-
\frac{1}{\z}\d_{W_i}(\l ) \psi_{W_j}(\z )\}+
$$
$$
+\frac{1}{\z}\e^{\t B}_A \D_{W_j}\,_{\t B}^C\d_{W_i}(\l )V_C
-\frac{1}{\l}\e^{\t B}_A \D_{W_i}\,_{\t B}^C\d_{W_j}(\z )V_C
. \eqno(38)
$$
{}From eqs.(36) one obtains the equations for the variation 
$\d_{W_i}(\l )\psi_{W_j}(\z )$:
$$
[\p_{\t A}+\z \e^B_{\t A}V_B, \d_{W_i}(\l )\psi_{W_j}(\z )]
=\z \e^B_{\t A}[\psi_{W_j}(\z ), \d_{W_i}(\l )V_B]+
(\D_{W_j}\,_{\t A}^B+\z \e^C_{\t A}\D_{W_j}\,_ C^B)\d_{W_i}(\l )V_B
.\eqno(39)$$
Using the identities
$$
\D_{X_{a}}\,_{\t A}^{B}=0,\quad 
\D_{X_{\hat a}}\,_{\t A}^{B}+ \z \e^C_{\t A}\D_{X_{\hat a}}\,_C^{B}=
(Z_{\hat a}^\z -\frac{\z}{2}\dot Z_{\hat a}^\z)\e^B_{\t A}
,\eqno(40)
$$
where $Z_{\hat a}^\z$ are the components of vector fields
$$
Z_{\hat a} = Z_{\hat a}^\z \p_\z , \ [Z_{\hat a}, Z_{\hat b}]=
f^{\hat c}_{\hat a \hat b}Z_{\hat c}
$$
$$
Z_{\hat 1}^\z =-\frac{i}{2}(1+\z^2),\
Z_{\hat 2}^\z =\frac{1}{2}(1-\z^2),\
Z_{\hat 3}^\z =i\z,\
\dot Z_{\hat a}^\z\equiv\frac{d}{d\z}Z_{\hat a}^\z ,
\eqno(41)
$$
we obtain the solution of eqs. (39) in the form
$$
\d_{W_i}(\l )\psi_{W_j}(\z )=\frac{\z}{(\l -\z )}\{\psi_{[W_i, W_j]}(\z )-
[\psi_{W_i}(\l ), \psi_{W_j}(\z )]-W_i^\z\p_\z\psi_{W_j}(\z )\}+
$$
$$
+\frac{\l}{(\l -\z )^2}W_j^\z\{\psi_{W_i}(\l )- \psi_{W_i}(\z )\}
,\eqno(42)
$$
where $W_a^\z :=0,\ W_{\hat a}^\z := Z_{\hat a}^\z.$

Substituting (42) into (38), we obtain the following expression
for the commutator of two successive infinitesimal transformations:
$$
[\d_{W_i}(\l ), \d_{W_j}(\z )] V_A=
\frac{1}{(\l -\z )}
\{{\l}\d_{[W_i, W_j]}(\l )-{\z}\d_{[W_i, W_j]}(\z )\}V_A+
$$
$$
+\frac{1}{(\l -\z )^2}\{
\frac{\z}{\l}W_i^\l (
{\z}\d_{W_j}(\z )-{\l}\d_{W_j}(\l ))
+\frac{\l}{\z}W_j^\z (
{\z}\d_{W_i}(\z )-{\l}\d_{W_i}(\l ))\}V_A+
$$
$$
+\frac{1}{\z}\e^{\t B}_A \D_{W_j}\,_{\t B}^C\d_{W_i}(\l )V_C
-\frac{1}{\l}\e^{\t B}_A \D_{W_i}\,_{\t B}^C\d_{W_j}(\z )V_C
+
$$
$$
+\frac{1}{(\l -\z )}\{W_i^\z \p_\z (
{\z}\d_{W_j}(\z ))+{W_j}^\l\p_\l (
{\l}\d_{W_i}(\l ))\}V_A
. \eqno(43)
$$
In order to rewrite (43) in terms of the generators $\d_{W_i}^n=(2\pi i)^{-1}
\oint_{C'}d\l \ \l^{-n-1}\d_{W_i}(\l )$, it is convenient to introduce
$Y_0, Y_\pm$ instead of $X_{\hat a}$:
$$
Y_0:=iX_{\hat 3},\ 
Y_+:=-iX_{\hat 1}+X_{\hat 2},\ 
Y_-:=-iX_{\hat 1}-X_{\hat 2},\ 
[Y_\pm , Y_0]=\pm Y_\pm ,\ [Y_+ , Y_-]=2 Y_0.
\eqno(44)
$$
Using (43) and (44), we obtain
$$
[\d^m_{X_a}, \d^n_{X_b}]=\d^{m+n}_{[X_a,X_b]},\ m,n,...\ge 0,
\eqno(45)
$$ 
$$
[\d^m_{Y_0}, \d^n_{Y_0}]=2(m-n)\d^{m+n}_{Y_0},\ 
[\d^m_{Y_+}, \d^n_{Y_+}]=2(m-n)\d^{m+n-1}_{Y_+},\
[\d^m_{Y_-}, \d^n_{Y_-}]=2(m-n)\d^{m+n+1}_{Y_-},
$$
$$
[\d^m_{Y_0}, \d^n_{Y_+}]=\d^{m+n}_{[Y_0,Y_+]}+2m\d^{m+n-1}_{Y_0}
-2n\d^{m+n}_{Y_+},
$$
$$
[\d^m_{Y_0}, \d^n_{Y_-}]=\d^{m+n}_{[Y_0,Y_-]}+2m\d^{m+n+1}_{Y_0}
-2n\d^{m+n}_{Y_-},\
$$
$$
[\d^m_{Y_+}, \d^n_{Y_-}]=\d^{m+n}_{[Y_+,Y_-]}+2m\d^{m+n+1}_{Y_+}
-2n\d^{m+n-1}_{Y_-},
\eqno(46)
$$
$$
[\d^m_{Y_0}, \d^n_{X_a}]=-2n\d^{m+n}_{X_a},\
[\d^m_{Y_+}, \d^n_{X_a}]=-2n\d^{m+n-1}_{X_a},\
[\d^m_{Y_-}, \d^n_{X_a}]=-2n\d^{m+n+1}_{X_a},
\eqno(47)
$$
{}Formulae (45) mean that $\{\d^m_{X_a}\}$ are the generators
of the affine Lie algebra $sl(2,C)\otimes C[\l ]$, which is the subalgebra
in $sl(2,C)\otimes C[\l , \l^{-1}].$ From (46) one can see 
that $\d^m_{Y_0}, \d^m_{Y_+}$ and $\d^m_{Y_-}$ generate three 
different Virasoro-like subalgebras of the symmetry algebra.

Thus, the new algebra of `hidden symmetries' of the SDG equations with 
generators\\ $\{\d^m_{X_1}, \d^m_{X_2}, \d^m_{X_3}, \d^m_{Y_0}, \d^m_{Y_+}, 
\d^m_{Y_-}\}$ forms a Kac-Moody-Virasoro algebra with 
commutation relations (45) -- (47). This algebra has the same commutation
relations as a subalgebra of the symmetry algebra
of the self-dual Yang-Mills equations [22].

\vspace{1cm}

{\bf 5. Commutators of symmetries and comments}

\vspace{0.2cm}

In Sec.4 and Sec.3 eqs.(36) on $\psi_{W_j},\ W_j\in so(4, C),$ and 
eqs.(21) on $\psi_M$, $M\in$ sdiff$(\Sigma^2)$, have been written out. From 
these equations one can derive the equations for the variations of 
the vector fields $\psi_{W_j}$ and $\psi_M$:
$$
[\p_{\t A}+\z\e^B_{\t A}V_B, \d_M(\l )\psi_{W_j}(\z )]=
\z\e^B_{\t A}[\psi_{W_j}(\z ), \d_M(\l )V_B]+
$$
$$ 
+(\D_{W_j}\,_{\t A}^B+\z\e^C_{\t A}\D_{W_j}\,_{C}^B)\d_M(\l )V_B,
\eqno(48a)
$$
$$
[\p_{\t A}+\l\e^B_{\t A}V_B, \d_{W_j}(\z )\psi_M(\l )]=
\l\e^B_{\t A}[\psi_M(\l ), \d_{W_j}(\z )V_B] 
.\eqno(48b)
$$
We have the following solutions of these equations:
$$
\d_M(\l )\psi_{W_j}(\z )=\frac{\z}{\l-\z}
[\psi_{W_j}(\z ), \psi_M(\l )]+\frac{\l}{(\l-\z)^2}W^\z_j
\{\psi_M(\l ) - \psi_M(\z )  \}
,\eqno(49a)
$$
$$
\d_{W_j}(\z )\psi_M(\l )=\frac{\l}{\l-\z}\{
[\psi_{W_j}(\z ), \psi_M(\l )]+W^\l_j\p_\l\psi_M(\l )  \}
.\eqno(49b)
$$
Then after some computation we obtain the following expression for the 
commutator
$$
[\d_M(\l ), \d_{W_j}(\z )]V_A=\frac{1}{(\l-\z)^2}\{\frac{\l}{\z}
W^\z_j(\z \d_M(\z )-\l\d_M(\l ))\}V_A +
$$
$$
+\frac{1}{\z}\e^{\t B}_{A}\D_{W_j}\,_{\t B}^C\d_M(\l )V_C + 
\frac{1}{(\l-\z )}W_j^\l\p_\l(\l\d_M(\l ))V_A.
\eqno(50)
$$ 
Using the definition of $\d_M^m,\ \d_{W_j}^n$, formulae (44) and 
the commutator (50), we obtain  
$$
[\d^m_{X_a}, \d^n_M]=0,\
[\d^m_{Y_0}, \d^n_M]=-2n\d^{m+n}_M,
\eqno(51a)
$$
$$
[\d^m_{Y_+}, \d^n_M]=-2n\d^{m+n-1}_M,\
[\d^m_{Y_-}, \d^n_M]=-2n\d^{m+n+1}_M,\
m,n,...\ge 0
.\eqno(51b)
$$ 
Thus, the `hidden symmetries' of the SDG equations (12) form the 
infinite-dimensional Lie algebra with the commutation relations (27),
(45)--(47) and (51).

Notice, that taking $\z =0$ in (49b), we obtain the action of 
$\d_{W_j}^0$ on $\psi_M (\l  )$:
$$
\d^0_{X_a}\psi_M(\l )=[\psi^0_{X_a}, \psi_M(\l )],\
\d^0_{X_{\hat a}}\psi_M(\l )=[\psi^0_{X_{\hat a}}, \psi_M(\l )]+
Z^\l_{\hat a}\p_\l\psi_M(\l )  
.\eqno(52)
$$
{}From (52) it follows that $\d^0_{X_a}$ acts on 
$\psi_M (\l  )$ as the Lie derivative along the vector field
$\psi^0_{X_a}$, and $\d^0_{X_{\hat a}}$ acts on $\psi_M (\l  )$ 
as the Lie derivative along the ``lifted" vector field 
$\psi^0_{X_{\hat a}}+Z_{\hat a}$ (cf. [31,32,22] for the SDYM case). 
The reason is that
the vector fields $\psi^0_{X_a}$, defined on the manifold $M^4$, have the
trivial lift on the twistor space $M^4\times B^2$  ($B^2\simeq S^2$ 
for Euclidean 
signature and $B^2\simeq H^2$ for the signature  (2, 2)), 
and the lift of the vector fields $\psi^0_{X_{\hat a}}$ is nontrivial.

Remember that $\d_{W_j}^0$ act on the metric as Lie derivatives: 
$\d_{W_j}^0g^{\mu\nu}={\cal L}_{\psi_{W_j}^0}g^{\mu\nu}$.
Therefore one can consider reductions of the SDG equations (12) and of the
linear system (21) for them by imposing the invariance conditions of the 
tetrad and of $\psi_M$ with respect to the vector fields 
$\{\psi_{W_j}^0\}$. For example, conditions 
$\d_{Y_0}^0V_A={\cal L}_{\psi_{Y_0}^0}V_A=0$
reduce the SDG equations to the $sl(\infty )$-Toda field equation
(see, e.g., [23,26]). Since  $\d_{Y_0}^0$ generates the Lie algebra
diff$(S^1)=\{\d_{Y_0}^n\}$ of the group Diff$(S^1)$, then the space 
of solutions of
the $sl(\infty )$-Toda field equation can be obtained from the space
$\cal M$ of solutions of the SDG equations by factorization under
the group Diff$(S^1)$. The imposing of $\d_{Y_0}^0$-symmetry
(from which there also follow the symmetries under 
$\d_{Y_0}^n, \ n\ge 1$),
automatically reduces the algebra of hidden symmetries of the SDG equations
to the well-known algebra $w_\infty\simeq$ sdiff$(\Sigma^2)$ of symmetries
of the $sl(\infty )$-Toda field equation [25]. Namely, only the subalgebra 
with generators $\{\d_M^0\}$ will preserve the symmetry condition (this 
algebra is a normalizer of the algebra diff$(S^1)$ in the symmetry algebra).

Analogously, the conditions $\d_{X_3}^0V_A={\cal L}_{\psi_{X_3}^0}V_A=0$ 
reduce the SDG equations to the Gibbons-Hawking equations [24,26  ],
describing, in particular, ALE gravitational instantons. From the 
symmetry with respect to $\d_{X_3}^0$
there follows the symmetry with respect to the whole algebra 
$\{\d_{X_3}^n\}$ of the 
Abelian loop group $LU(1)=C^\infty(S^1,U(1))$. 
Therefore, the space of solutions of
the Gibbons-Hawking equations is obtained from the space
$\cal M$ of solutions of the SDG equations by factorization under 
the group $LU(1)$. From the commutation relations (45)--(47) and (51) 
it follows that 
the subalgebra with generators $\{\d_M^n, \d_{Y_0}^n, \d_{Y_+}^n, 
\d_{Y_-}^n, \ n\ge 0\}$ will preserve the 
symmetry condition. This algebra has not been described 
in the literature before. 

\newpage

{\bf References}

\begin{enumerate}
\item E.Kiritsis, C.Kounnas and D.L\"ust, Int. J. Mod. Phys. {\bf A9}
      (1994) 1361;  
      M.Bianchi, F.Fucito, G.Rossi and M.Martellini, Nucl. Phys. {\bf B440}
      (1995) 129; 
      M.J.Duff, R.R.Khuri and J.X.Lu, Phys. Rep. {\bf 259} (1995) 213.
\item H.Ooguri and C.Vafa, Mod. Phys. Lett. {\bf A5} (1990) 1389;
      Nucl. Phys. {\bf B361} (1991) 469; Nucl. Phys.  {\bf B451} (1995) 121. 
\item N.Berkovits and C.Vafa, Nucl. Phys.  {\bf B433} (1995) 123;
      N.Berkovits, Phys. Lett.  {\bf B350} (1995) 28; Nucl. Phys.  {\bf B450} 
      (1995) 90. 
\item C.M.Hull, String Dynamics at Strong Coupling, hep-th/9512181;
      C.Vafa, Evidence for F-Theory, hep-th/9602022;
      D.Kutasov and E.Martinec, New Principle for String/ Mem\-brane 
      Unification, hep-th/9602049.
\item A.Font, L.Ib\'a\~{n}ez, D.L\"ust and F.Quevedo, Phys. Lett. 
      {\bf B249} (1990) 35; A.Sen, Int. J. Mod. Phys.  {\bf A9} (1994) 3707;
      A.Giveon, M.Porrati and E.Rabinovici, Phys. Rep.  {\bf 244} (1994) 77;
      E.Alvarez, L.Alvarez-Gaum\'e and Y.Losano, Nucl. Phys.  (Proc.  Sup. ) 
      {\bf 41} (1995) 1;
      C.M.Hull and P.K.Townsend, Nucl. Phys.  {\bf B438} (1995) 109; 
      Nucl. Phys.  {\bf B451} (1995) 525; J.H.Schwarz, Superstring 
      Dualities, hep-th/9509148.
\item S.W.Hawking and G.W.Gibbons, Phys. Rev.  {\bf 13} (1977) 2752; 
      G.W.Gibbons, M.J.Perry and S.W.Hawking, Nucl. Phys.  {\bf B138} (1978) 
      141;  
      G.W.Gibbons and M.J.Perry, Nucl. Phys.  {\bf B146} (1978) 90. 
\item K.Yamagishi and G.F.Chapline, Class. Quantum Grav. {\bf 8} (1991) 427;
      K.Yamagishi, Phys. Lett. {\bf B259} (1991) 436.
\item R.Penrose, Gen. Rel. Grav. {\bf 7} (1976) 31.
\item M.F.Atiyah, N.J.Hitchin and I.M.Singer, Proc. R. Soc. Lond. {\bf A362}
      (1978) 425.
\item K.P.Tod and R.S.Ward, Proc. R. Soc. Lond. {\bf A386} (1979) 411;
      N.J.Hitchin, Math. Proc. Camb. Phil. Soc. {\bf 85} (1979) 465;
      R.S.Ward, Commun. Math. Phys. {\bf 78} (1980) 1.
\item M.Ko, M.Ludvigsen, E.T.Newman and K.P.Tod, Phys. Rep. {\bf 71} 
      (1981) 51.
\item C.P.Boyer and J.F.Pleba\'nski, J. Math. Phys. {\bf 18} (1977) 1022;
      J. Math. Phys. {\bf 26} (1985) 229;
      C.P.Boyer, Lect. Notes Phys. Vol.189 (1983) 25.
\item J.F.Pleba\'nski, J. Math. Phys. {\bf 16} (1975) 2395.
\item C.P.Boyer and P.Winternitz, J. Math. Phys. {\bf 30} (1989) 1081.
\item K.Takasaki, J. Math. Phys. {\bf 30} (1989) 1515; J. Math. Phys. 
      {\bf 31} (1990) 1877; Preprint RIMS-747, 1991.
\item Q-Han Park, Phys. Lett. {\bf B238} (1990) 287; Phys. Lett. 
     {\bf B257} (1991) 105;  J.Hoppe and Q-Han Park, Phys. Lett. {\bf B321}
     (1994) 333.
\item J.D.E.Grant, Phys. Rev. {\bf D48} (1993) 2606; 
      I.A.B.Strachan, J. Math. Phys. {\bf 36} (1995) 3566.
\item V.Husain, Class. Quantum Grav. {\bf 11} (1994) 927;  
      J. Math. Phys. {\bf 36} (1995) 6897.
\item A.Ashtekar, T.Jacobson and L.Smolin, Commun. Math. Phys. {\bf 115}
      (1988) 631; 
      L.J.Mason and E.T.Newman, Commun. Math. Phys. {\bf 121} (1989) 659; 
      R.S.Ward, Class. Quantum Grav. {\bf 7} (1990) L217.
\item S.Chakravarty, L.Mason and E.T.Newman, J. Math. Phys. {\bf 32} (1991)
      1458; R.S.Ward, J. Geom. Phys. {\bf 8} (1992) 317;
      C.Castro, J. Math. Phys. {\bf 34} (1993) 681;
      V.Husain, Phys. Rev. Lett. {\bf 72} (1994) 800;
      J.F.Pleba\'nski and M.Przanowski, Phys. Lett. {\bf A212} (1996) 22.
\item L.J.Mason and N.M.J.Woodhouse, {\it Integrability, Self-Duality
      and Twistor Theory}, Clarendon Press, Oxford, 1996.
\item A.D.Popov and C.R.Preitschopf, Phys. Lett. {\bf B374} (1996) 71.
\item C.Boyer and J.Finley, J. Math. Phys. {\bf 23} (1982) 1126;
      J.Gegenberg and A.Das, Gen. Rel. Grav. {\bf 16} (1984) 817.
\item G.W.Gibbons and S.W.Hawking, Phys. Lett. {\bf 78B} (1978) 430;
      Commun. Math. Phys. {\bf 66} (1979) 291.
\item I.Bakas, In: Proc. of the Trieste Conf. ``Supermembranes and Physics
      in 2+1 Dimensions", eds. M.Duff, C.Pope and E.Sezgin, World 
      Scientific, Singapore, 1990, p.352; 
      Q-Han Park, Phys. Lett. {\bf B236} (1990) 429;
      I.Bakas, Commun. Math. Phys. {\bf 134} (1990) 487;
      K.Takasaki and T.Takebe, Lett. Math. Phys. {\bf 23} (1991) 205.
\item I.Bakas,  Phys. Lett. {\bf B343} (1995) 103;
      I.Bakas and K.Sfetsos,  Phys. Lett. {\bf B349} (1995) 448;
      E.Alvarez, L.Alvarez-Gaum\'e and I.Bakas, Nucl. Phys. 
      {\bf B457} (1995) 3;
      Supersymmetry and Dualities, hep-th/9510028.
\item E.Bergshoeff, R.Kallosh and T.Ortin, Phys. Rev. {\bf D51} (1995) 3003;
      S.F.Hassan, Nucl. Phys. {\bf B460} (1996) 362;
      K.Sfetsos, Nucl. Phys. {\bf B463} (1996) 33.
\item M.L\"uscher and K.Pohlmeyer, Nucl. Phys. {\bf B137} (1978) 46;
      E.Brezin, C.Itzykson, J.Zinn-Justin and J.-B.Zuber, Phys. Lett.
      {\bf 82B} (1979) 442; H.J. de Vega, Phys. Lett. {\bf 87B} (1979) 233;
      H.Eichenherr and M.Forger,  Nucl. Phys. {\bf B155} (1979) 381;
      L.Dolan, Phys. Rev. Lett. {\bf 47} (1981) 1371; Phys. Rep. {\bf 109} 
      (1984) 3;
      K.Ueno and Y.Nakamura, Phys. Lett. {\bf 117B} (1982) 208;
      Y.-S.Wu, Nucl. Phys. {\bf B211} (1983) 160;
      L.-L.Chau, Lect. Notes Phys. Vol. 189 (1983) 111.
\item J.H.Schwarz, Nucl. Phys. {\bf B447} (1995) 137; Nucl. Phys. {\bf B454}
      (1995) 427.
\item A.C.Newell, {\it Solitons in Mathematics and Physics}, SIAM, 
      Philadelphia, 1985. 
\item M.Legar\'e and A.D.Popov, Phys. Lett. {\bf A198} (1995) 195; 
      T.A.Ivanova and A.D.Popov, Theor. Math. Phys. {\bf 102} (1995) 280;
      JETP Lett. {\bf 61} (1995) 150.
\item T.A.Ivanova and A.D.Popov, Phys. Lett. {\bf A205} (1995) 158;
      Phys. Lett. {\bf A170} (1992) 293; M.Legar\'e, J.Nonlinear Math.Phys. 
      {\bf 3} (1996) 266.

\end{enumerate}

\end{document}